\documentclass[preprint]{aastex}

\begin{document}

\title{Superwind-driven Intense H$_2$ Emission in NGC 6240 II:
Detailed Comparison of Kinematical and Morphological Structures
of the Warm and Cold Molecular Gas}

\author{Youichi Ohyama,}
\affil{Subaru Telescope, National Astronomical Observatory of Japan,
650 N. A`ohoku Place, University Park, Hilo, HI 96720}

\author{Michitoshi Yoshida,}
\affil{Okayama Astrophysical Observatory, National Astronomical
Observatory
    of Japan, Kamogata-cho, Asakuchi-gun, Okayama 719-0232, Japan}

\author{and Tadafumi Takata}
\affil{Subaru Telescope, National Astronomical Observatory of Japan,
650 N. A`ohoku Place, University Park, Hilo, HI 96720}

\begin{abstract}
We report on our new analysis of the spatial and kinematical distribution
of warm and cold molecular gas in the nearby, prototypical, luminous
infrared galaxy NGC 6240, which was undertaken to explore the origin of its
unusually luminous H$_2$ emission. The gas components are known to be
distributed between the two merging nuclei, forming an off-nuclear
molecular gas concentration. By comparing three-dimensional emission-line
data (in space and velocity) of CO (J=$2-1$) in the radio and H$_2$ in
the near infrared, we are able to search for the spatial and kinematical
conditions under which efficient H$_2$ emission is produced in much
more detail than has previously been possible. In particular, we focus
on the H$_2$ emitting efficiency, defined in terms of the intensity
ratio of H$_2$ to CO [$I$(H$_2$)/$I$(CO)], as a function of velocity.
We derive this by utilizing the recent high-resolution
three-dimensional data presented by Tecza et al. (2000). The
integrated H$_2$ emitting efficiency is calculated by integrating the
velocity profile of H$_2$ emitting efficiency in blue, red, and total
(blue + red) velocity regions of the profile. We find that (1) both
the total H$_2$ emitting efficiency and the blue-to-red ratio of the
efficiency are larger in regions surrounding the CO and H$_2$
intensity peaks, and (2) the H$_2$ emitting efficiency and the
kinematical conditions in the warm molecular gas are closely related
to each other. We compare two possible models that might explain these
characteristics: a large-scale collision between the molecular gas
concentration and the merging nuclei, and a collision between the
molecular gas concentration and the external superwind outflow from
the southern nucleus. The latter model seems more plausible, since it
can reproduce the enhanced emitting efficiency of blueshifted H$_2$
around the molecular gas concentration, if we assume that the
superwind blows from the southern nucleus toward us, hitting the entire
gas concentration from behind. In this model, internal cloud-cloud
collisions within the molecular gas concentration are enhanced by the
interaction with the superwind outflow, and efficient and intense
shock-excited H$_2$ emission is expected as a result of the
cloud-crushing mechanism. The observed spatial distribution of the
H$_2$ emitting efficiency can be explained if there is a greater
kinematical disturbance in the outer part of the molecular gas
concentration, as a result of the interaction with the superwind
outflow, and also more frequent cloud-cloud collisions in the region.
In addition, the kinematical influence of the superwind on the
molecular gas concentration should be larger at bluer velocities, and
the collision frequency is expected to be larger at bluer velocities,
explaining the relationship between velocity and the H$_2$ emitting
efficiency.

\end{abstract}


\section{INTRODUCTION}

NGC 6240 is a nearby\footnote{In this paper, the distance is
assumed to be 98 Mpc, following, for consistency, Heckman, Armus, \&
Miley (1987) and Ohyama et al. (2000).} luminous infrared galaxy
(e.g., Wright, Joseph \& Meikle 1984; Joseph \& Wright 1984),
comprising two merging nuclei (north and south nuclei: hereafter, N
and S nuclei, respectively) separated by 1.9\arcsec~ (Condon et al.
1982; Fried \& Schulz 1983; Eales et al. 1990; Thronson et al. 1990;
Herbst et al. 1990; Colbert, Wilson, \& Bland-Hawthorn 1994; Sugai et
al. 1997; Tacconi et al. 1999; Ohyama et al. 2000; Scoville et al.
2000; Tecza et al. 2000; Beswick et al. 2001), and is well known for
its unusually luminous H$_2$ emission in the near infrared (e.g.,
Rieke et al. 1985; DePoy, Becklin, \& Wynn-Williams 1986). Although
many excitation mechanisms for the H$_2$ have been proposed to date
[shock excitation (Rieke et al. 1985; DePoy et al. 1986; Lester,
Harvey, \& Carr 1988; Elston \& Maloney 1990; Herbst et al. 1990; van
der Werf et al. 1993; Sugai et al. 1997); UV fluorescence (Tanaka,
Hasegawa, \& Gatley 1991); X-ray heating (Draine \& Woods 1990); and
formation pumping (Mouri \& Taniguchi 1995)], H$_2$ line ratio
analyses have revealed that shock excitation is responsible for almost
all the H$_2$ emission (Sugai et al. 1997; Ohyama et al. 2000; Tecza
et al. 2000). However, because of its huge H$_2$ luminosity and
exceptionally large H$_2$/Br $\gamma$ intensity ratio, the origin of
NGC 6240's H$_2$ emission has been the subject of considerable debate;
star formation activity is not strong enough to power all the observed
H$_2$ luminosity (e.g., Rieke et al. 1985; Draine \& Woods 1990).
Interestingly, most of the H$_2$ emission comes from the region
between the nuclei, although both Br $\gamma$ and [Fe II] (originating
from OB type stars and supernovae in star-forming regions) are
detected from each nucleus (Herbst et al. 1990; van der Werf et al.
1993; Sugai et al. 1997; Ohyama et al. 2000; Tecza et al. 2000). Two
possible models have been proposed to explain this intense,
off-nuclear, shock-excited H$_2$ emission. In the first, the emission
is caused by shock heating at the interface between the two colliding
galaxies (Tecza et al. 2000 and references
therein), while in the second it is due to the interaction
between the hot gas associated with the superwind activity of the S
nucleus and the tidally-produced molecular gas concentration between
the two nuclei (Ohyama al. 2000).

Recent high-resolution, sensitive CO observations of NGC 6240 at radio
wavelengths have revealed that a significant fraction of the system's
molecular gas is concentrated not at the two nuclei but rather
occupies the region between them, forming a massive thick disk with
highly turbulent motion (Tacconi et al. 1999). Hereafter we refer to
the CO gas between the nuclei as the off-nuclear molecular gas or
molecular gas concentration. A violent tidal disturbance during the
merging process of the two nuclei could be responsible for the
formation of such an off-nuclear molecular gas disk, although the
creation of such a massive concentration of gas away from the merging
nuclei has no theoretical explanation (e.g., Barnes 2002 and
references therein; see also Tecza et al. 2002). The peak of the H$_2$
emission is located between the S nucleus and the CO peak (Ohyama et
al. 2000; Tecza et al. 2000).

The first step in exploring the reason
for the source's unusually intense H$_2$ emission was the analysis of
the velocity field and the excitation conditions of the H$_2$ lines
(Ohyama et al. 2000; Tecza et al. 2000 and references therein). In
this paper we take the second step by conducting a detailed comparison
of the morphological and kinematical conditions of H$_2$ and CO lines.
This work is motivated by the fact that H$_2$ and CO
are molecular species with different excitation conditions;
comparisons of their emission lines should give us some new insights on the
conditions under which high H$_2$ emitting efficiency (H$_2$ flux
per molecular gas content) is likely. Such a comparison has only
recently become possible, as a result of the availability of sub-arcsecond, sensitive mapping observations of both CO
and H$_2$. Tecza et al. (2000) conducted three-dimensional H$_2$
mapping observations of this galaxy under sub-arcsec seeing conditions,
while Tacconi et al. (1999) carried out high-resolution CO ($J=2-1$)
mapping observations with a sub-arcsec beam. Both data have been re-sampled
to make a direct comparison between the cold and warm molecular gas at
the same scale (Tecza et al. 2000). Based on these data, we discuss how
H$_2$ emission can be excited efficiently, focusing on the H$_2$/CO
intensity ratio as an indicator of the H$_2$ emitting efficiency.

\section{Data and Analysis}

All data that we analyze in this paper were reproduced from Figure 3 of
Tecza et al. (2000), in which grids of both the CO ($J=2-1$) and
H$_2$ velocity profiles were presented. These profiles were reproduced
from the original data of both Tecza et al. (2000) and Tacconi et al.
(1999) for H$_2$ and CO, respectively. The velocity resolution of the
original H$_2$ data was 150 km s$^{-1}$, while the spatial resolution
was 0.8\arcsec~ - 1.0\arcsec~, the seeing conditions at the time of
observation (Tecza et al. 2000). In the CO observations, the full
width at half maximum (FWHM) of the elliptical Gaussian was beam
0.9\arcsec $\times$ 0.5\arcsec~, while the velocity resolution was 6.5
km s$^{-1}$. These data were re-sampled onto $0.6 \times 0.6$
arcsec$^2$ sub-regions ($290 \times 290$ pc$^2$) around the double
nuclei over $3.6 \times 4.2$ arcsec$^2$ ($1.7 \times 2.0$ kpc$^2$)
regions, and re-sampled at 140 km s$^{-1}$ velocity resolution (Tecza
et al. 2000). These data enabled us to compare the CO and H$_2$
velocity profiles directly, with a resolution that was sufficient to separate
the two merging nuclei and the surrounding regions.

The velocity profiles were integrated over blue and red velocity
regions ($dV \equiv V-V_{\rm sys}= -300 \sim 0$ km s$^{-1}$ and $dV =0
\sim 300$ km s$^{-1}$ for blue and red, respectively, where $V_{\rm
sys}$ is the systemic velocity) in each sub-region to measure the blue
and red intensities of CO and H$_2$ [$I_{\rm blue}$(CO), $I_{\rm
red}$(CO), $I_{\rm blue}$(H$_2$), and $I_{\rm red}$(H$_2$)]. The
velocity cut-offs at $dV = \pm 300$ km s$^{-1}$ were applied in the
integration because it is necessary to measure the intensities in the
same velocity range for all sub-regions in order to make a fair
comparison of their spatial distribution. We found that some
sub-regions had a weak CO intensity at larger velocity offsets from
the systemic velocity, so that the intensity could not be accurately measured
out to the velocity cut-off. We eliminated the
integrated intensity data of the sub-regions from the following
analyses if the normalized intensity of CO (normalized to be unity at
the CO peak position: see Figure 3 of Tecza et al. (2000) and Figure 2
of this paper) was 0.05 (the typical noise level) or less at the
cut-off velocity. The cut-off velocity ($dV = \pm 300$ km s$^{-1}$)
was chosen as a compromise: although a larger cut-off velocity is
desirable for studying the properties of the high-velocity gas
components, applying a larger ($dV > 300$ or $<-300$ km s$^{-1}$)
velocity cut-off would discard a larger number of sub-regions due
to a lack of useful integrated intensity data. Note that since the
observed line widths ($\sim 1000$ km s$^{-1}$ and $\sim 1600$ km
s$^{-1}$ for CO and H$_2$ at full width at zero intensity,
respectively: Tacconi et al. 1999; Tecza et al. 2000) are wider than
the applied velocity cut-off (at $dV = \pm 300$ km s$^{-1}$), about
19\% and 33\% of the total CO and H$_2$ fluxes fall outside the
velocity range for the integration, and were not counted in the
analysis that follows. Single Gaussian profile fitting was used to
measure the velocities of both lines [$V$(CO) and $V$(H$_2$)],
although some areas show more complicated emission line profiles, such
as a double-peaked profile (Tecza et al. 2000). Figure 1 shows the
total (blue+red) integrated intensity and the velocity maps of both CO
and H$_2$ [$I_{\rm blue}$(CO)$+I_{\rm red}$(CO), $I_{\rm
blue}$(H$_2$)$+I_{\rm red}$(H$_2$), $V$(CO), and $V$(H$_2$)]. The
peaks of both lines are located between the two nuclei and slightly
east of the S nucleus ($\sim 0.6$\arcsec)
\footnote{ Note that the higher resolution H$_2$ map presented by Tecza et
al. (2000) and other previous H$_2$ maps (e.g., Sugai et al. 1997)
showed the H$_2$ peak between the double nuclei, although our Figure 1
shows the peak at a slightly different position. This is an artefact that was
probably caused by the re-sampling of the data onto the $0.6\arcsec
\times 0.6\arcsec$~ sub-regions adopted in producing Figure 3 of Tecza
et al. (2000) and our Figure 1. A similar artefact is also seen in the
map of CO. The adopted velocity range for integrating the profiles
(from $dV = -300$ km s$^{-1}$ to $dV = 300$ km s$^{-1}$) may also
affect the positions of the CO and H$_2$ peaks in our maps. }. The
velocities of both lines change along PA$\sim 45$ deg, in the sense
that they are blueshifted around the S nucleus and are redshifted in
both the NE and SW regions. Simple models of the velocity field, such
as a rotation or a pure expansion/contraction, have difficulty
reproducing these observations.

To compare the properties of CO and H$_2$ emissions, we here
introduce a new index, the H$_2$ emitting efficiency.
The efficiency is defined as the intensity ratio of H$_2$ and CO
[$I$(H$_2$)/$I$(CO)], and was calculated over the entire velocity range on
each sub-region (Figure 2).
Note, however, that these ratios are shown only at velocities where the
normalized CO intensity is larger than 0.05 (the typical noise level) to avoid
obtaining erroneous ratios in regions or velocities where CO is weak.
Next, the integrated H$_2$ emitting efficiencies for blue
[$R_{\rm blue}\equiv \int I({\rm H_2})/I({\rm CO}) dV/\Delta V$,
  integrating over the range $dV = -300$ to $0$ km s$^{-1}$], for red ($R_{\rm red}$,
integrating over the range $dV = 0$ to $+300$ km s$^{-1}$), and for blue plus
red (total) ($R_{\rm total}$, integrating over the range $dV = -300$ to $+300$
km s$^{-1}$) were calculated, in order to examine the spatial and
velocity distributions of the efficiencies.
Here $\Delta V$ is the velocity range for the integration, i.e.,
$\Delta V=300$ km s$^{-1}$ for blue- and red-, and
$\Delta V=600$ km s$^{-1}$ for total-integrated H$_2$ emitting efficiencies.
A velocity cut-off (at $dV = \pm 300$ km s$^{-1}$) was applied in
integrating the H$_2$ emitting efficiency, in the same way as for the
calculation of the integrated intensities.
Also, the integrated H$_2$ emitting efficiencies in sub-regions where the
normalized intensity of CO at the cut-off velocity was less than 0.05 were not
used in the following analysis.
The blue-to-red ratio of the H$_2$ emitting efficiency
($R_{\rm blue}$/$R_{\rm red}$) could then be calculated.
These values ($R_{\rm total}$ and $R_{\rm blue}$/$R_{\rm red}$) are mapped
together with the velocity difference between CO and H$_2$
[$\Delta V_{\rm H_2-CO}\equiv V$(H$_2$)$-V$(CO)] in Figure 3.

Figure 3 illustrates our main results: (1) The total H$_2$ emitting
efficiency ($R_{\rm total}$) is smaller around the CO peak, and
becomes higher with distance from the peak; (2) the blue-to-red ratio
of the H$_2$ emitting efficiency ($R_{\rm blue}$/$R_{\rm red}$) is
smaller around the CO peak, and becomes larger with distance from the
peak; and (3) the velocity difference ($\Delta V_{\rm H_2-CO}$) is
smaller around the CO peak, while the H$_2$ emission becomes more
blueshifted with respect to CO with distance from the peak. All these
trends are seen almost systematically as a function of position, and
are evident well over our resolution ($\sim 0.6\arcsec$ in space and
140 km s$^{-1}$ in velocity). It seems noteworthy that the peak of
H$_2$ emitting efficiency corresponds neither to the peak
position of CO nor to that of H$_2$.

We searched for correlations between the measured quantities ($R_{\rm
total}$ and $R_{\rm blue}$/$R_{\rm red}$); the results are shown in
Figures 4 and 5. In these plots, data are marked with different
symbols for different groups of sub-regions, which were divided
according to their spatial positions around the CO intensity peak: the
central region with higher H$_2$ and CO intensities between the two
nuclei (nuclear region), the region northeast of the central region
(NE region), and the southwestern region (SW region). Two other
sub-regions, the most southwesterly regions where H$_2$ intensity
increases with distance from the nucleus (the SW sub-group) and the
most northeasterly regions with double-peaked profiles (the NE
sub-group), are grouped separately. These sub-regions are also
indicated in Figure 2. We found a tight negative correlation between
$I$(CO) and $R_{\rm total}$ (Figure 4). A Spearman rank-order test for
the correlation between log[$R_{\rm total}$] and log[$I$(CO)] showed
that the probability $P$ of consistency with the null hypothesis
(i.e., no correlation between the two quantities) is less than 0.001
for any sub-grouping (all the regions, all the regions except the SW
sub-group, and all regions except for both SW and NE sub-groups).
Therefore we consider that we have definitely shown that the total
H$_2$ emitting efficiency is enhanced in the outer regions with weaker
CO emission. Two other positive correlations were found, between $\Delta
V_{\rm H_2-CO}$, $R_{\rm blue}$/$R_{\rm red}$, and $R_{\rm total}$
(Figure 5). Although these correlations look less tight, a Spearman
rank order test showed that the probability $P$ of the null hypothesis
is 0.20, 0.02, and 0.04 for the above three sub-groupings,
respectively, in $\Delta V_{\rm H_2-CO}$ vs. $R_{\rm blue}$/$R_{\rm
red}$ diagram, and 0.02, 0.001, and $<$0.001 for the same three
sub-groupings, respectively, in $\Delta V_{\rm H_2-CO}$ vs. $R_{\rm
total}$ diagram, indicating the presence of significant correlations
between these quantities at greater than the 98\% significance level
if we exclude the SW sub-group.

\section{DISCUSSION}

\subsection{H$_2$ Excitation Mechanism}

Our analysis has shown that the H$_2$ emitting efficiency and the
kinematical conditions of the warm gas are closely related.
Such a correlation is a good indication of shock heating, in
which bulk kinetic energy is converted to energy radiated through the
H$_2$ line. Although our analyses were carried out only for fluxes
within the velocity range of $dV = -300 \sim 300$ km s$^{-1}$, it
seems reasonable to interpret the correlation as an indication of the
major mechanism of the excitation of H$_2$, because this velocity
range contains 81\% and 67\% of the total fluxes of CO and H$_2$,
respectively (see Section 2 above). In some inner sub-regions where we
can trace the H$_2$ emitting efficiency at even bluer velocities ($dV
< -300$ km s$^{-1}$), we found that the efficiency is generally higher
at bluer velocity ($dV < -300$ km s$^{-1}$) than at lower velocity
($dV = -300 \sim 0$ km s$^{-1}$), and this fact supports a model in
which shock excitation of H$_2$ is a major excitation mechanism around
NGC 6240. Although other H$_2$ excitation mechanisms (e.g., UV
fluorescence, X-ray heating, and formation pumping) cannot be
rejected, their contribution to the total must be less important.
We note that the same conclusion has previously been reached
based on a line-ratio analysis of the near-infrared spectra (Sugai
et al. 1997; Tecza et al. 2000; and Ohyama et al. 2000).

\subsection{The Superwind Model}

Since correlations between the H$_2$ emitting efficiency and the
kinematical conditions of the molecular gas were found not only on the
SW but also on the NE side of the CO peak, on scales of $2\arcsec$, the
molecular gas concentration seems likely to be surrounded by gas that
has a simple global motion. The motion is probably directed
predominantly toward the observer, and the interaction between the gas
that has this blueshifted motion and the molecular gas concentration
would increase the H$_2$ emitting efficiency on the bluer velocity side
of the CO emission. The question then arises: what is the origin of
this blueshifted gas motion? In the following, we discuss two possible
explanations.

One possibility is that the merging nuclei are interacting with the
off-nuclear molecular gas as they rotate around each other. Although
there are no good theoretical explanations for the formation of such
an off-nuclear molecular gas concentration in merging galaxies
(e.g., Barnes 2002 and references therein; see also Tecza et al.
2002), here we assume that such concentration has formed as a result
of tidal effects during the merging process. Since the H$_2$ emitting
efficiency is relatively higher at the SW portion of the
concentration, a collision between the S galaxy and the concentration
might contribute to the excitation of H$_2$ emission. The extinction
maps (Scoville et al. 2000; Tecza et al. 2000) have revealed the
presence of a dusty region at 0\arcsec -- 1.0\arcsec\ NNE of the
S nucleus, and the position of the maximum extinction ($\sim$
0.2\arcsec -- 0.3\arcsec~ NNE of the S nucleus) coincides spatially
with the CO peak. This indicates that the S galaxy is behind the dusty
molecular gas concentration from our point of view. Since the S
nucleus is known to be moving toward us around the concentration
(Fried \& Ulrich 1985; Tecza et al. 2000), a violent collision between
the two would enhance the blueshifted H$_2$ emission at the SW side of
the CO peak. If this is the case, however, the redshifted H$_2$
emission, rather than the blueshifted component, would be enhanced on
the opposite side of the CO peak, because the N galaxy is rotating
away from us on the opposite side of the concentration. This model
therefore cannot explain the fact that blue-enhanced H$_2$ emitting
efficiency is observed on both sides of the CO peak.

The alternative model is that an outflow associated with the superwind
activity of the S nucleus interacts with the off-nuclear molecular gas
concentration. The superwind activity of this galaxy has been
recognized as a result of its large-scale ($\sim 20$ kpc)
shock-excited filamentary nebulae (e.g., Armus, Heckman, \& Miley
1990; Keel et al. 1990) and the expanding motion of the H$_2$ emitting
clouds around the nuclear region (Ohyama et al. 2000). Since both of
the individual galaxies are known to show flat rotation curves in
their stellar velocity fields (Tecza et al. 2000), disk structures are
likely to be present around each nucleus, even though the system is an
advanced merger. Therefore we expect any superwind outflow to exhibit
a bi-directional conical shape pointing toward the disk polar
direction (e.g., Tomisaka \& Ikeuchi 1988; Heckman et al. 1990;
Suchkov et al. 1994; Strickland \& Stevens 2000), although the shape
of the outflow might be distorted due to the inhomogeneous
distribution of the medium around the nucleus. An $H-$band image of
the S nucleus reveals an elongated structure in the NW-SE direction,
suggesting the presence of a disk whose minor axis is directed NE-SW
at an inclination angle of $i\simeq 60$ deg (Scoville et al. 2000;
Tecza et al. 2000). Also, as mentioned earlier, the S nucleus seems to
be located behind the molecular gas concentration. Assuming that the
off-nuclear molecular gas concentration formed during the merging
process of the two nuclei, before the superwind started to blow, we
would expect that, if a bi-conical superwind outflow emanated from the
S nucleus, it would interact with the gas concentration from the rear.
This interaction would produce blueshifted H$_2$ emission around the
CO peak (i.e., not only at the SW but also at the NE sides of the CO peak)
if the opening angle of the outflow were wide enough to cover the
whole molecular gas concentration. The oppositely directed cone of the
superwind, pointing away from us, would not contribute significantly
to the H$_2$ luminosity, as it would not interact with the off-nuclear
molecular gas. Since this model has no major difficulties in
reproducing the observations, we regard it as more plausible than the
model in which the S nucleus is colliding with the molecular gas
concentration. A schematic picture of the superwind model is shown in
Figure 6. Note that the real situation would be much more complicated
due to the complicated medium distribution around the merging nuclei.

Following Heckman et al. (1993), clouds accelerated by the superwind
outflow will have a terminal velocity of $v_{\rm cloud} = 400
L^{1/2}_{{\rm bol}, ~11} r^{-1/2}_{\rm kpc} N^{-1/2}_{21}$ km
s$^{-1}$, where $L_{\rm bol, ~11}$ is the bolometric luminosity in
units of $10^{11}$ $L_{\rm \sun}$, $r_{\rm kpc}$ is the initial
distance of the cloud from the nucleus, and $N_{21}$ is the column
density of the interstellar clouds in units of $10^{21}$ cm$^{-2}$.
Adopting $L_{\rm bol, ~11}=4.6$ (Sanders \& Mirabel 1995), $r \sim
1.0$ kpc (the projected distance between the S nucleus and the CO peak
multiplied by a de-projection factor of $\sim 2$), and $N_{21} = 1 -
20$\footnote{ Here we adopt $N_{21}=1$ as a minimum acceptable value
corresponding to the typical value for the Milky Way. We also adopt
$N_{21}\sim 20$ as a maximum acceptable value corresponding to one
tenth of the total column density inferred from molecular gas
observations of $N_{21}\gtrsim 200$ (Tacconi et al. 1999). } , this
equation implies that the clouds would be blown off at a speed of
between a few and several hundreds of km s$^{-1}$, which is almost
comparable to the observed line width of H$_2$ ($\sim 500$ km s$^{-1}$
in FWHM: e.g., Ohyama et al. 2000; Tecza et al. 2000) and CO ($\sim
260 - 400$ km s$^{-1}$ in FWHM: Tacconi et al. 1999). Since the column
density is likely to decline outward around the CO peak, falling off
in a similar way to the CO surface brightness, (Figure 1a)
\footnote{The total column density of the molecular gas integrated
over the
line of sight is proportional to the observed integrated CO intensity,
so the total CO intensity map is similar
to the map of the total column density.
Although the column density of each cloud may not be
proportional to the total one, they are likely to be almost
proportional to each other if we assume a simple homogeneous
distribution of small clouds within the molecular gas
concentration.},
i.e., the column density of each cloud is larger near the CO peak
and is smaller at the outer region, so that clouds in the outer part
of the
molecular gas concentration would be accelerated up to faster speeds
than those near to the CO peak.
Therefore the observed trend in the relatively blueshifted H$_2$ emission with
respect to the CO emission in the outer part of the molecular gas concentration
can be explained naturally by the model of superwind-molecular gas
concentration interaction.
This fact further suggests that kinematical disturbance due to the superwind
also raises the H$_2$ emitting efficiency within the molecular gas
concentration, given the positive correlation between the H$_2$ emitting
efficiency and $\Delta V_{\rm H_2-CO}$.

It has been pointed out that a C-type shock ($v_{\rm shock} \lesssim
40$ km s$^{-1}$), rather than a J-type shock with faster shock
velocity, is responsible for exciting most of the H$_2$ emission,
although most of the emission comes from clouds within a faster global
velocity field ($dV > 100$ km s$^{-1}$) (Sugai et al. 1997). One way
to resolve this apparent contradiction is to suggest that internal
cloud-cloud collisions within the clumpy medium, at speeds of $\sim
30$--$50$ km s$^{-1}$, excite the H$_2$ emission, while the clouds move
along the global velocity field as a whole (Tecza et al. 2000). Since
the molecular gas concentration is known to form a turbulent thick
rotating disk (Tacconi et al. 1999), it is likely that the
concentration is actually composed of numerous smaller clouds (Tecza
et al. 2000). A high-resolution velocity dispersion map of the CO
(Figure 4 of Tacconi et al. 1999) reveals the presence of regions with
higher velocity dispersion just around the CO peak (especially to the
SE and NW sides of the CO peak). Therefore it seems very likely that
the interaction of small clouds within the molecular gas
concentration with the external superwind provides an enhanced random
velocity at the outer part of the CO concentration, as well as the
blueshift of the H$_2$ with respect to the CO.

\subsection{The Cloud-Crushing Model}

One of the merits of utilizing cloud-cloud collisions within the molecular gas
concentration to give rise to the H$_2$ emission is that a higher efficiency of
energy deposition into the molecular gas clouds from the supernova explosions
can be expected as a result of the cloud-crushing mechanism
(Cowie, McKee, \& Ostriker 1981), and the observed huge H$_2$ luminosity can
be explained (Draine \& Woods 1990; Elston \& Maloney 1990).
This therefore raises the question of whether the observed trends in H$_2$ emitting
efficiency, as well as the total H$_2$ luminosity, can be explained in
terms of the cloud-crushing mechanism.
In this model, we may consider, as a simple case, that some fraction of the kinetic
energy of the small clouds, randomly moving within the bulk motion of the
rotating molecular gas concentration, is dissipated through internal shocks
as a result of cloud-cloud collisions, and gives rise to H$_2$ emission.
Then, the H$_2$ surface brightness at a velocity of $v$ can be expressed as
$I_{\rm H_2} (v) = 0.5 \Sigma_{\rm H_2} (v) \zeta (v) <v_{\rm internal}^2>
\epsilon / \Delta t$, where $\Sigma_{\rm H_2} (v)$ is the surface mass density
of the molecular gas concentration whose velocity is $v$, $\zeta (v)$ is the
mass fraction of the small clouds that give rise to H$_2$ emission through
the cloud-crushing mechanism at a velocity $v$, $<v_{\rm internal}^2>$ is the
mean of the squared internal random velocity of the small clouds,
$\epsilon$ is the fraction of the energy dissipated in the shock which is
radiated in the H$_2$ 1-0 $S$(1) line,
and $\Delta t$ is the timescale of the cloud-cloud collisions.
Since the surface mass density of the clouds $\Sigma_{\rm H_2} (v)$ is
proportional to the observed CO surface brightness [$I_{\rm CO} (v)$] at the
velocity of $v$, the H$_2$ emitting efficiency can be expressed as
$I_{\rm H_2} (v)/I_{\rm CO} (v) \propto \zeta (v) <v_{\rm internal}^2>
\epsilon / \Delta t$, by definition.
Since we are considering a situation in which the cloud-crushing mechanism is
responsible for {\it all} the H$_2$ emission, we may substitute a constant $\epsilon$
($\simeq 0.3$) and a constant internal velocity
($<v_{\rm internal}^2>^{1/2} = 40 - 50$ km s$^{-1}$) into the
equation, representing an extreme case of the cloud-crushing mechanism working at its
maximum efficiency (Draine \& Woods 1990).
Then, the H$_2$ emitting efficiency can be expressed as
$I_{\rm H_2} (v)/I_{\rm CO} (v) \propto \zeta (v) / \Delta t$.
This expression indicates that the H$_2$ emitting efficiency is higher when
(1) a larger mass fraction of the clouds is involved in the production
of H$_2$ emission through
the cloud-crushing mechanism, and/or (2) the collision frequency is higher
(i.e., the collision timescale is shorter).
The observed close relationship between the H$_2$ emitting efficiency and the
kinematical conditions in the molecular gas clouds can be explained by
this expression, if the cloud-crushing mechanism is efficient:
The outer part of the off-nuclear molecular-gas concentration is
likely to be more strongly influenced kinematically by the interaction
with the external superwind (see Section 3.2 above),
and the H$_2$ emitting efficiency is
expected to be higher, since a larger mass fraction of the clouds will
produce H$_2$ as a result of internal cloud-cloud shocks and a higher
frequency of the cloud-cloud collisions. In addition, the efficiency
is expected to be higher at bluer velocities, because these are likely to
be caused as a result of a more violent kinematical influence from the
superwind. Therefore, the cloud-crushing mechanism within a molecular
gas concentration disturbed by the interaction with the external
superwind outflow seems to be able to reproduce the observed trends of
the H$_2$ emitting efficiency in both space and velocity, as well as
the higher-than-normal H$_2$ luminosity. Although the mechanism by
which the off-nuclear molecular gas concentration between the two
merging nuclei is created is not clear, the presence of such a gas
concentration is necessary if the model is to reproduce the observed
intense and efficient H$_2$ emission. We look forward to future
theoretical studies of the origin of such a gas concentration.

\acknowledgments

We thank Matthias Tecza who kindly allowed us to see his doctoral
thesis and gave us useful comments on the ``Data and Analysis''
section of this paper.


\figcaption{
Maps of the total integrated intensities and velocities of CO and H$_2$.
North is up, and east is to the left.
The positions of the N and S nuclei and the peak position of CO are marked in
all plots.
Regions where valid data are not available are left blank in these maps.
See the main text for the data selection criteria.
(a) A map of total integrated CO intensity distribution.
The map is shown in a linear scale with contours at an interval of
an arbitrary constant.
(b) Same as (a), but for H$_2$.
The map is shown in a linear scale with contours at an interval of
an arbitrary constant.
(c) A map of CO velocity.
Red and blue colors indicate redshifted and blueshifted velocities.
The map is shown in a linear scale with contours at constant intervals of
40 km s$^{-1}$.
Small numbers within the map indicate the velocity with respect
to the systemic velocity in units of km s$^{-1}$.
(d) Same as (c), but for H$_2$.
The contours are drawn at constant intervals of 40 km s$^{-1}$.
}

\figcaption{
Velocity profiles of CO (black), H$_2$
(blue), and H$_2$/CO intensity ratio (red) at various positions.
Each plot shows the profiles integrated over 0.6\arcsec~ by
0.6\arcsec~ sub-regions, taken from Figure 3 of Tecza et al. (2000).
The intensity profiles of both CO and H$_2$ are normalized at the position
of the CO peak.
The scale for both CO and H$_2$ profiles is shown on the left side, and
that for H$_2$/CO intensity ratio profile is shown on the right side.
The regions are grouped into five sub-regions, and are shown with
different symbols:
the central region with higher H$_2$ and CO intensities around the two
nuclei (double circles),
the northeastern region from the central region (filled squares),
the southwestern region (open squares),
the most southwesterly region, where H$_2$ intensity increases with
distance from the nuclei (upper triangle),
and the most northeasterly region, with double-peaked profiles (lower
triangle).
Regions without any marks are not used in the correlation analyses in
Figures 4 and 5, where a CO velocity profile is not available out to
the velocity cut-off at $\pm 300$ km s$^{-1}$.
See the main text for a discussion of the velocity cut-off.
}

\figcaption{
Maps of the H$_2$ emitting efficiencies and the velocity difference between
H$_2$ and CO.
North is up, and east is to the left.
The positions of the N and S nuclei and the CO peak are marked in all plots.
(a) Total integrated CO intensity distribution.
This map is the same as Figure 1 (a), but is shown here again for ease of
comparison with other maps.
(b) A map of the velocity difference between CO and H$_2$
($\Delta V_{\rm H_2-CO}$).
The blue color indicates the region where H$_2$ is relatively blueshifted with
respect to CO.
Small numbers within the map indicate $\Delta V_{\rm H_2-CO}$ in units of
km s$^{-1}$.
Contours are drawn at constant intervals of 20 km s$^{-1}$.
(c) A map of the total H$_2$ emitting efficiency ($R_{\rm total}$).
The green color indicates regions with larger $R_{\rm total}$.
Small numbers within the map indicate $R_{\rm total}$.
The minimum contour drawn near the S nucleus is drawn at
$R_{\rm total} = 1.25$, and other contours are drawn at constant intervals of
$\delta R_{\rm total} = 0.5$.
(d) A map of the blue-to-red ratio of the H$_2$ emitting efficiency ($R_{\rm
blue}/R_{\rm red}$).
The blue color indicates the region with relatively enhanced $R_{\rm blue}$
over $R_{\rm red}$.
Small numbers within the map indicate $R_{\rm blue}/R_{\rm red}$.
The minimum contour drawn between the two nuclei is drawn at
$R_{\rm blue}/R_{\rm red} = 2.0$, and other contours are drawn at constant
intervals of $\delta R_{\rm blue}/R_{\rm red} = 0.2$.
}

\figcaption{The correlation between integrated CO intensity
[$I$(CO)] and total H$_2$ emitting efficiency ($R_{\rm total}$).
The units of $I$(CO) are arbitrary.
Both axes are shown on a logarithmic scale.
Data are plotted with different symbols according to the sub-region
from which they come,
as shown in Figure 2.
}

\figcaption{Plots of the blue-to-red ratio of the H$_2$ emitting efficiency
($R_{\rm blue}/R_{\rm red}$) and the velocity difference between CO and H$_2$
($\Delta V_{\rm H_2-CO}$) as a function of the total H$_2$ emitting efficiency
($R_{\rm total}$).
The symbols used are the same as in Figures 2 and 4.
}

\figcaption{A proposed schematic picture of the superwind model of NGC 6240
seen from our point of view (left) and from the side (right).
The two galactic nuclei are rotating around the off-nuclear molecular gas
concentration.
One of the bi-directional cones of the superwind from the S nucleus
(blue) points towards us and interacts with the off-nuclear molecular gas
concentration.
The interaction between the superwind and the concentration
causes the cloud-crushing mechanism to work efficiently in the region
between the S nucleus and the concentration (green).
Another cone flowing in the opposite direction from the S nucleus (red)
would not give rise to such efficient H$_2$ emission, as there is no off-nuclear
molecular gas on this side.
See the main text for a more detailed discussion of the model.
}

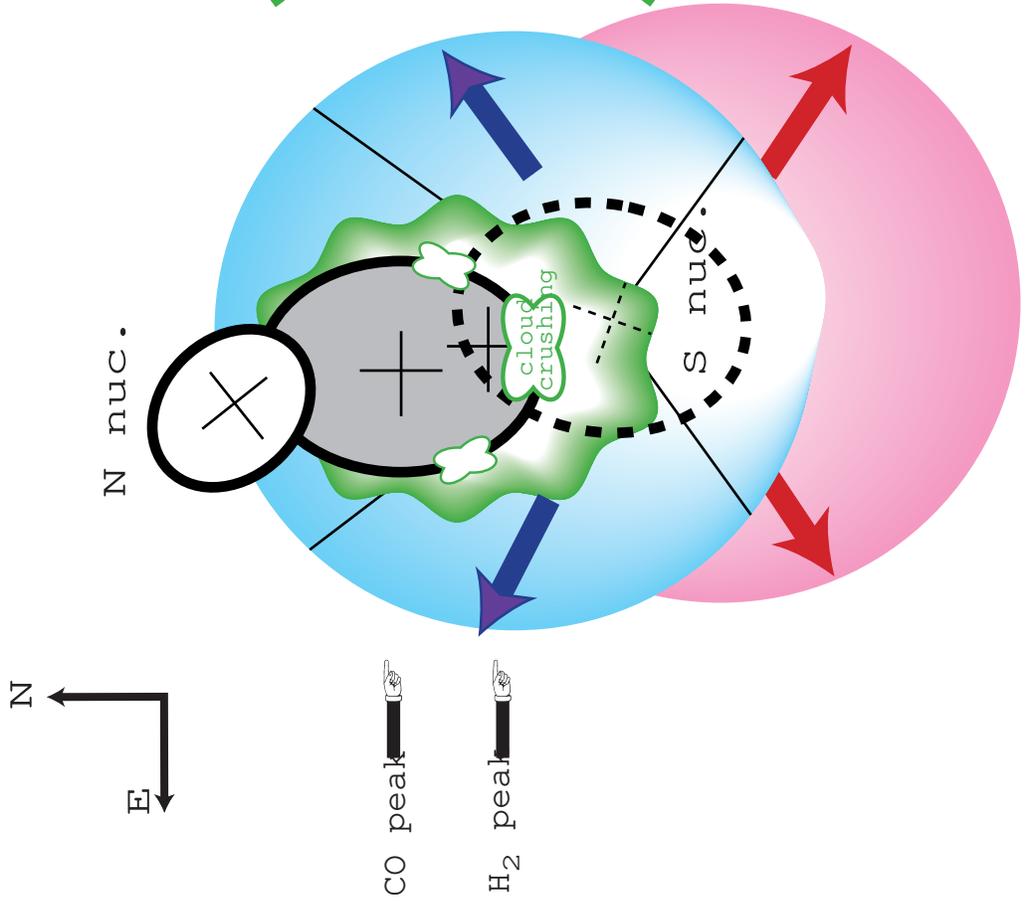
\end{document}